\documentclass[iop,apjl]{emulateapj}

\usepackage{graphicx}  
\usepackage{dcolumn}   
\usepackage{bm}        
\usepackage{amsfonts,amsmath,amssymb,mathrsfs}
\usepackage{color}
\usepackage{epsfig}
\usepackage{natbib,times}
\usepackage{url}
\usepackage{times}
\allowdisplaybreaks[4]

\pdfoutput=1

\def\VEC#1{\mbox{\boldmath $#1$}}

\newcommand{\f}[2]{\frac{#1}{#2}}

\makeatletter
\newcommand{\vast}{\bBigg@{4}}
\newcommand{\Vast}{\bBigg@{5}}
\makeatother

\DeclareMathAlphabet{\mathpzc}{OT1}{pzc}{m}{it}

\shorttitle{Relativistic magnetohydrodynamics with Landau-Lifshitz radiation reaction}
\shortauthors{Wenshuai Liu}

\begin{document}

\title{One-fluid relativistic magnetohydrodynamics equations for a two-fluid plasma with the Landau-Lifshitz radiation reaction force}

\author{Wenshuai {Liu}\altaffilmark{1}$^{,a}$, Weihao {Bian}\altaffilmark{1}$^{,b}$, Bixuan {Zhao}\altaffilmark{1}, Liming {Yu}\altaffilmark{1}, Chan {Wang}\altaffilmark{1}}

\altaffiltext{1}{Department of Physics and Institute of Theoretical Physics, Nanjing Normal University, Nanjing 210023, China}
\altaffiltext{a}{674602871@qq.com}
\altaffiltext{b}{whbian@njnu.edu.cn}

\begin{abstract}
By taking into account the radiation reaction force, we derive a set of one-fluid relativistic magnetohydrodynamics (RMHD) equations with the Landau-Lifshitz radiation reaction force based on a relativistic two-fluid plasma. These equations could be used to situations where spatiotemporal scales of plasma's motion are sufficiently large.
\end{abstract}

\keywords{accretion, accretion disks --- radiation reaction ---
    magnetohydrodynamics (MHD)}

\section{Introduction}
 Recent observations show that there is a strong magnetic field with several hundred gauss around the supermassive black hole residing in the center of the Milky Way \citep{1}. The magnetic field around a black hole can be produced by dynamo mechanism by an accretion disk made of plasma orbiting a black hole \citep{2,27,28}. It's clear that a neutron star can have huge surface magnetic field up to $10^{14}$ G \citep{21,22,23,24,25,26}, and many neutron stars are found in a binary system in which the neutron star may accrete material from the companion star \citep{29}. The presence of such a magnetic field can affect a test charged particle's motion significantly. Then the magnetic field could cause drastic effect on the evolution of an accretion disk around a neutron star \citep{30,31} or a stellar-mass black hole accreting material from its companion star in a binary system \citep{50} , or around a supermassive black hole in AGN \citep{51}. Results from both magnetohydrodynamics and general relativistic magnetohydrodynamics simulations show that magnetized stellar-mass black hole or supermassive black hole in AGN accretes material from the accretion disk due to magnetorotational instability \citep{38,51}, and accretion materials flow onto neutron star along the magnetic field line \citep{30,31,32,33,34,35}. In these high energy accretion disks, it is suggested that the disks are made of ion-electron plasma \citep{3}. The plasma disk is often described by relativistic magnetohydrodynamics (RMHD) without the effects of Landau-Lifshitz radiation reaction which may play an important role when the spatiotemporal scales of plasma's motion are sufficiently large. Besides the method of particle-in-cell \citep{4,5,7,6,8,10,9} which incorporates the radiation reaction forces, applications of radiation reaction to the fluid dynamics perspective have been studied by \cite{11}, \cite{12} and \cite{13}. Here with the one-fluid description using a relativistic two-fluid approximation of plasma consisting of positively and negatively charged particles, we derive one-fluid relativistic magnetohydrodynamics of two-fluid plasma in which Landau-Lifshitz radiation reaction is included.

\section{Equations derived}
In this work, the background spacetime is assumed to be Minkowski spacetime defined by $ds^2=-(dx^0)^2+(dx^1)^2+(dx^2)^2+(dx^3)^2=\eta_{\mu\nu}dx^\mu dx^\nu$ where $(x^0,x^1,x^2,x^3)=(t,x,y,z)$. We set $c=1$, $\epsilon_0=1$, $\mu_0=1$ which represent the speed of light, the dielectric constant, and the magnetic permeability in vacuum are all unity.

To derive the relativistic one-fluid equations of the two-fluid plasma with radiation reaction, we define the average and difference variables which is same to \cite{14}. The two-fluid plasma consists
of positively charged particles, each has charge $e$ and mass $m_+$,
and of negatively charged particles, each with charge $-e$ and mass $m_-$. Here we list the 4-velocity and 4-current density as follows
\begin{eqnarray}
U^\mu &=& \frac{1}{\rho} ( m_+ n_+ u_+^\mu + m_- n_- u_-^\mu )
\label{1} \\
J^\mu &=& e(n_+ u_+^\mu - n_- u_-^\mu)
\label{2}
\end{eqnarray}
where $m=m_+ + m_-$ and $\rho=m_+ n_+ + m_- n_-$

The generalized RMHD equations for the two-fluid plasma with the Maxwell equations are given
as follows \cite{14}
\begin{eqnarray}
\partial_\nu (\rho U^\nu) &=& 0  \label{r1} \\
\partial_\nu  \left [
h \left (U^\mu U^\nu + \frac{\mu}{q^2} J^\mu J^\nu \right ) \right ]
&=& -\partial^\mu p + J^\nu {F^\mu}_\nu   \label{r2} \\
\frac{1}{q}   \partial_\nu  \left [
\frac{\mu h}{q} (U^\mu J^\nu + J^\mu U^\nu )
- \frac{2\mu \Delta \mu h}{q^2} J^\mu J^\nu \right ] \nonumber
&=& \frac{1}{2q} \partial^\mu (\Delta \mu p - \Delta p)
+\left ( U^\nu - \frac{\Delta \mu}{q} J^\nu \right) {F^\mu}_\nu
\nonumber \\
&-&\eta [J^\mu
+Q (1+\Theta) U^\mu]\\
\partial_\nu \hspace{0.3em} ^*F^{\mu\nu} &=& 0  \label{r3} \\
\partial_\nu F^{\mu\nu} & = & J^\mu  \label{r4}
\label{rmhdohm}
\end{eqnarray}
where $\mu = m_+ m_-/m^2$, $\Delta \mu = (m_+ -m_-)/m$, $p=p_+ + p_-$, $\Delta p=p_+ - p_-$, $q=ne$, $n=\frac{\rho}{m}$, $F_{\mu\nu}$ is the electromagnetic field tensor, $\hspace{0.3em} ^*F^{\mu\nu}$ is the dual tensor density of $F_{\mu\nu}$, $\mu, \nu=0,1,2,3$, $Q=U^\nu J_\nu$ and $\Theta$ the thermal
energy exchange rate from negatively charged fluid to the positive fluid (see Appendix A of \cite{14} for details).

Before deriving the main equations, we begin with the dynamical equation for a charged particle of the mass $m$ with the electric charge $e$ in a background magnetic field. This dynamical equation reads
\begin{equation}
m\frac{du^{\mu }}{ds}=eF^{\mu \nu }u_{\nu }+g^{\mu }
\label{eq01a}
\end{equation}
where $u_{\mu }$ is the four-velocity and $\mu =0,1,2,3$

The radiation reaction force in the Lorentz-Abraham-Dirac form is given by

 \begin{equation}
g^{\mu }=\frac{2e^{2}}{3}\left[ \frac{d^{2}u^{\mu }}{ds^{2}}-u^{\mu }u^{\nu }\frac{d^{2}u_{\nu }}{ds^{2}} \right]
\label{eq02}
\end{equation}

We follow Landau and Lifshitz (LL) \citep{15} assuming that $g^\mu$ is small compared with Lorentz force in the instantaneous rest frame of the charged particle, allowing us to express equation (\ref{eq02}) as follows in detail
\begin{eqnarray}
g^{\mu }&=&\frac{2e^{3}}{3m}\left\{ \frac{\partial F^{\mu \nu }}{
\partial x^{\lambda }}u_{\nu }u_{\lambda }
-\frac{e}{m}\left[ F^{\mu
\lambda }F_{\nu \lambda }u^{\nu }-\left( F_{\nu \lambda }u^{\lambda }\right)
\left( F^{\nu \kappa }u_{\kappa }\right) u^{\mu }\right] \right\}\label{eq10}
\end{eqnarray}
where the first term is negligible compared to other terms. The reason is that the force, called Frenkel force \citep{20}, acting on a charged particle with a spin of freedom moving in an external electromagnetic field, the case of a plane wave for example, is quite larger than the first term in equation (\ref{eq10}). In the regime where the classical equation of motion is adopted without Frenkel force from quantum effects, therefore, it is consistent to neglect the first term in equation (\ref{eq10}) \citep{9,16}. Then equation (\ref{eq01a}) becomes
\begin{eqnarray}
m\frac{du^{\mu }}{ds} & = & eF^{\mu \nu }u_{\nu }
-\frac{2e^{4}}{3m^{2}}\left[ F^{\mu
 \lambda }F_{\nu \lambda }u^{\nu }-\left( F_{\nu \lambda }u^{\lambda }\right)
 \left( F^{\nu \kappa }u_{\kappa }\right) u^{\mu }\right]
\end{eqnarray}

When inserting radiation reaction into generalized RMHD, we should get the total radiation reaction force on the two species of the two-fluid plasma. To do so we must get $u^{\mu}_\pm$ and $n_\pm$,  that is to say, ten unknowns. Equation (\ref{1}) and (\ref{2}) are only eight equations. To solve this, we can combine two additional equations as follows
\begin{equation}
\gamma_\pm = u^0_\pm
\label{3}
\end{equation}
where $\gamma_\pm=\sqrt{\frac{1}{1-((\frac{u^1_\pm}{u^0_\pm})^2+(\frac{u^2_\pm}{u^0_\pm})^2+(\frac{u^3_\pm}{u^0_\pm})^2)}}$

Using equation (\ref{1}), (\ref{2}), we get
\begin{equation}
n_\pm u_\pm ^\mu = \frac{1}{m} \left  (
\rho U^\mu \pm \frac{m_\mp}{e} J^\mu \right )
\end{equation}
 combined with (\ref{3}), we could get $u^{\mu}_\pm$ and $n_\pm$ as follows
\begin{eqnarray}
n_\pm & = & \frac{\sqrt{\left  (\rho U^0 \pm \frac{m_\mp}{e} J^0 \right )^2- \sum_{n=1}^3\left(\rho U^n \pm \frac{m_\mp}{e} J^n \right )^2}}{m}\label{17}\\
u^\mu_\pm & = & \frac{\left  (\rho U^\mu \pm \frac{m_\mp}{e} J^\mu \right )}{\sqrt{\left  (\rho U^0 \pm \frac{m_\mp}{e} J^0 \right )^2- \sum_{n=1}^3\left(\rho U^n \pm \frac{m_\mp}{e} J^n \right )^2}}\label{18}
\end{eqnarray}

Then the radiation reaction force of the two species in the fluid element are
\begin{equation}
f^{\mu}_{RR\pm}=-\frac{2n_\pm e^{4}}{3m^{2}_\pm}\left[ F^{\mu
 \lambda }F_{\nu \lambda }u^{\nu }_\pm-\left( F_{\nu \lambda }u^{\lambda }_\pm\right)
 \left( F^{\nu \kappa }u_{\pm \kappa }\right) u^{\mu }_\pm\right]
 \label{eq9}
\end{equation}

Inserting fluid element's radiation reaction equation (\ref{eq9}) into momentum density equation (\ref{r2}), we get the momentum density equation with radiation reaction as
\begin{eqnarray}
\partial_\nu  \left[ h \left (U^\mu U^\nu + \frac{\mu}{q^2} J^\mu J^\nu \right ) \right] & = & -\partial^\mu p + J^\nu {F^\mu}_\nu
 \nonumber \\
& - & \frac{2n_+ e^{4}}{3m^{2}_+}\left[ F^{\mu
 \lambda }F_{\nu \lambda }u^{\nu }_+
 - \left( F_{\nu \lambda }u^{\lambda }_+\right)
 \left( F^{\nu \kappa }u_{+ \kappa }\right) u^{\mu }_+\right]
 \nonumber \\
& - & \frac{2n_- e^{4}}{3m^{2}_-}\left[ F^{\mu
 \lambda }F_{\nu \lambda }u^{\nu }_-
 - \left( F_{\nu \lambda }u^{\lambda }_-\right)
 \left( F^{\nu \kappa }u_{- \kappa }\right) u^{\mu }_-\right] \label{r17}
\end{eqnarray}

The generalized RMHD equations for the ion-electron
plasma with only electron's radiation reaction (ion's radiation reaction is negligible due to the relatively large $m_+$) are given as follows with the limit $m \approx m_+ \gg m_-$ and an assumption $\Theta =0$
\begin{eqnarray}
\partial_\nu (\rho U^\nu) &=& 0  \label{eq777}\\
\partial_\nu  ( h U^\mu U^\nu )
&=& -\partial^\mu p + J^\nu {F^\mu}_\nu
- \frac{2n_- e^{4}}{3m^{2}_-}\left[ F^{\mu
 \lambda }F_{\nu \lambda }u^{\nu }_-
- \left( F_{\nu \lambda }u^{\lambda }_-\right)
 \left( F^{\nu \kappa }u_{- \kappa }\right) u^{\mu }_-\right] \label{eq8}\\
\frac{m_-}{mq}   \partial_\nu  \left [
\frac{h}{q} (U^\mu J^\nu + J^\mu U^\nu ) \right ]
&=& \frac{1}{q} \partial^\mu p_-
 +  \left ( U^\nu - \frac{1}{q} J^\nu \right) {F^\mu}_\nu
-  \eta [J^\mu + Q U^\mu] \label{eq501}\\
\partial_\nu^*F^{\mu\nu} &=& 0 \\
\partial_\nu F^{\mu\nu} &=& J^\mu \label{eq499}
\end{eqnarray}

The generalized RMHD equations for a pair plasma with radiation reaction are given
by $m_+ = m_- = m_{\rm e}$ ($m_{\rm e}$ is the mass of
electron) as
\begin{eqnarray}
\partial_\nu (\rho U^\nu) &=& 0  \label{11}   \\
\partial_\nu  \left [h
\left (U^\mu U^\nu + \frac{1}{(2q)^2} J^\mu J^\nu \right ) \right ]
&=& -\partial^\mu p + J^\nu {F^\mu}_\nu
 \nonumber \\
& - & \frac{2n_+ e^{4}}{3m^{2}_+}\left[ F^{\mu
 \lambda }F_{\nu \lambda }u^{\nu }_+
 - \left( F_{\nu \lambda }u^{\lambda }_+\right)
 \left( F^{\nu \kappa }u_{+ \kappa }\right) u^{\mu }_+\right]
 \nonumber \\
& - & \frac{2n_- e^{4}}{3m^{2}_-}\left[ F^{\mu
 \lambda }F_{\nu \lambda }u^{\nu }_-
 - \left( F_{\nu \lambda }u^{\lambda }_-\right)
 \left( F^{\nu \kappa }u_{- \kappa }\right) u^{\mu }_-\right]\label{22}\\
\frac{1}{4q}   \partial_\nu  \left [
\frac{h}{q} (U^\mu J^\nu + J^\mu U^\nu ) \right ]
&=& - \frac{1}{2q} \partial^\mu \Delta p + U^\nu {F^\mu}_\nu
- \eta [J^\mu + Q (1+\Theta) U^\mu]\label{33}\\
\partial_\nu^*F^{\mu\nu} &=& 0 \label{44}\\
\partial_\nu F^{\mu\nu} &=& J^\mu\label{55}
\label{pairrmhdohm}
\end{eqnarray}

When the inertia of current density, thermal electromotive force, and the Hall effect are ignored in equation (\ref{eq501}), equations (\ref{eq777})--(\ref{eq499}) change to the standard RMHD equations with radiation reaction
\begin{eqnarray}
\partial_\nu (\rho U^\nu) &=& 0  \label{eq12}\\
\partial_\nu  ( h U^\mu U^\nu ) &=& -\partial^\mu p + J^\nu {F^\mu}_\nu
-  \frac{2n_- e^{4}}{3m^{2}_-}\left[ F^{\mu
 \lambda }F_{\nu \lambda }u^{\nu }_-
 - \left( F_{\nu \lambda }u^{\lambda }_-\right)
 \left( F^{\nu \kappa }u_{- \kappa }\right) u^{\mu }_-\right] \label{eq13} \\
U^\nu {F^\mu}_\nu   &=& \eta [J^\mu + Q U^\mu]\label{eq14}\\
\partial_\nu^*F^{\mu\nu} &=& 0 \label{eq15}\\
\partial_\nu F^{\mu\nu} &=& J^\mu\label{eq16}
\end{eqnarray}

The next step we will rewrite equations (\ref{eq12})--(\ref{eq16}) in a form
of explicit time evolution conservation equations. Before rewriting, we investigate the approximate ratio of the Landau-Lifshitz radiation reaction force and the Lorentz force.

The Landau-Lifshitz radiation reaction for an electron in equation (\ref{eq10}) is valid when \citep{17}
\begin{equation}
\frac{\gamma_{\rm e}B}{B_{\rm c}}\ll 1
\label{b}
\end{equation}
that is to say, the magnetic field in the electron's rest frame shouldn't not exceed the classical critical magnetic field strength $B_{\rm c}=m^2_{e}c^4/e^3\approx 6\times 10^{15}~$G above which the Larmor radius of an electron in the electron's rest frame $m_{\rm e}c^2/eB$ becomes smaller than $r_{\rm e}$. But quantum effects could be important at even smaller magnetic field $B_{\rm QED}=m^2_{\rm e}c^3/\hbar e=\alpha_{\rm F}B_{\rm c}\approx 4.4\times 10^{13}~$G. Thus equation (\ref{b}) should be updated to \citep{17}
\begin{equation}
\frac{\gamma_{\rm e}B}{B_{\rm QED}}\ll 1
\label{f}
\end{equation}

In the four-dimensional form equation (\ref{eq10}) can be written as
\begin{eqnarray}
g^0&=&\frac{2e^{4}}{3m^{2}}u^0 \mathbf{E} \cdot (\mathbf{E}+\mathbf{v} \times \mathbf{B})
-  \frac{2e^{4}}{3m^{2}\displaystyle{\left( 1-{v^{2}}\right) }}u^0
\left\{ \left( \mathbf{E}+\mathbf{v\times B}\right) ^{2}-%
\left( \mathbf{v\cdot E}\right) ^{2}\right\} \label{eq18}
\end{eqnarray}
\begin{eqnarray}
\mathbf{g} &=&\frac{2e^{4}}{3m^{2}}u^0\left\{ \mathbf{E\times B}+
\left( \mathbf{B}\times \left( \mathbf{B\times v}\right) \right) +
\mathbf{E}\left( \mathbf{v\cdot E}\right) \right\}
-\frac{2e^{4}}{3m^{2}\displaystyle{\left( 1-{v^{2}}\right) }}
\mathbf{u}\left\{ \left( \mathbf{E}+\mathbf{v\times B}\right) ^{2}-\left( \mathbf{v\cdot E}\right) ^{2}\right\} \label{eq17}
\end{eqnarray}
where $i=1,2,3$, $\mathbf{v}=(v^1, v^2,v^3)$, $v^i=\frac{u^i}{u^0}$.

Thus, the ratio of the radiation reaction force and the Lorentz force acting on a electron is \citep{17}
\begin{equation}
\frac{|\mathbf{F_{\rm RR}}|}{|\mathbf{F_{\rm L}}|}\sim \frac{\gamma_{\rm e}^2 B}{B_{\rm c}}=\alpha_{\rm F}\gamma_{\rm e}\frac{\gamma_{\rm e}B}{B_{\rm QED}}
\label{c}
\end{equation}
Taking the accretion of rotating gas onto a newly born stellar mass black hole in Gamma-ray burst for instance, the magnetic field can reach up to $10^{15}$ G and $\gamma\sim10^2-10^3$. When we take $B=10^{8}$ G and $\gamma=10^3$, then $\frac{\gamma_{\rm e}B}{B_{\rm QED}}\approx0.00227$ from equation (\ref{f}), but $\frac{|\mathbf{F_{\rm RR}}|}{|\mathbf{F_{\rm L}}|}\sim \frac{\gamma_{\rm e}^2 B}{B_{\rm c}}\approx0.016$ from equation (\ref{c}). When we take $B=10^{10}$ G and $\gamma=10^3$, then $\frac{\gamma_{\rm e}B}{B_{\rm QED}}\approx0.227$ and $\frac{|\mathbf{F_{\rm RR}}|}{|\mathbf{F_{\rm L}}|}\sim \frac{\gamma_{\rm e}^2 B}{B_{\rm c}}\approx1.6$. From the above, the radiation reaction force acting on an electron in plasma could have significant effect on the dynamics of the electron, further having an significant effect on the current density in plasma. There exists situations for an electron with ultrarelativistic velocity that the Landau-Lifshitz radiation reaction force could be comparable with the Lorentz force in the laboratory frame, while being much
smaller in the electron's instantaneous rest frame \citep{15}.

Expanding equations (\ref{r1}), (\ref{r17}) and equations (\ref{r3})--(\ref{r4}) and including Landau-Lifshitz radiation reaction force acting on arbitrary masses of positively charged and negatively charged particles, $m_+$ and $m_-$ combined with  equation (\ref{eq18}) and (\ref{eq17}) and assuming ideal RMHD regime, we get the following one-fluid RMHD equations for two-fluid plasma

\begin{eqnarray}
\frac{\partial (\gamma{\rho})}{\partial t}+\nabla \cdot (\gamma\rho\mathbf{v}) &=& 0 \label{eq38}\\
\frac{\partial \mathbf{R}}{\partial t}+\nabla \cdot \mathbf{F} &=&
\frac{2n_-e^{4}}{3m_-^{2}}u_-^0\left\{ \mathbf{E\times B}
+\left( \mathbf{B}\times \left( \mathbf{B}\times\mathbf{v_-}\right)\right)+
\mathbf{E}\left( \mathbf{v_-\cdot E}\right) \right\}
 \nonumber \\
&-&  \frac{2n_-e^{4}}{3m_-^{2}\displaystyle{\left( 1-{v_-^{2}}\right) }}
\mathbf{u_-}\left\{ \left( \mathbf{E}
+\mathbf{v_-\times B}\right) ^{2}-
\left( \mathbf{v_-\cdot E}\right) ^{2}\right\}
\nonumber \\
&+&\frac{2n_+e^{4}}{3m_+^{2}}u_+^0\left\{ \mathbf{E\times B}
+\left( \mathbf{B}\times \left( \mathbf{B}\times\mathbf{v_+}\right)\right)+
\mathbf{E}\left( \mathbf{v_+\cdot E}\right) \right\}
 \nonumber \\
&-&  \frac{2n_+e^{4}}{3m_+^{2}\displaystyle{\left( 1-{v_+^{2}}\right) }}
\mathbf{u_+}\left\{ \left( \mathbf{E}
+\mathbf{v_+\times B}\right) ^{2}-
\left( \mathbf{v_+\cdot E}\right) ^{2}\right\} \label{39}\\
\frac{\partial \varepsilon}{\partial t}+\nabla \cdot \mathbf{R} &=&
\frac{2n_-e^{4}}{3m_-^{2}} u_-^0\mathbf{E} \cdot (\mathbf{E}+\mathbf{v_-} \times \mathbf{B})
 \nonumber \\
 &-&  \frac{2n_-e^{4}}{3m_-^{2}\displaystyle{\left( 1-{v_-^{2}}\right) }}u_-^0
\left\{ \left( \mathbf{E}+\mathbf{v_-\times B}\right) ^{2}-%
\left( \mathbf{v_-\cdot E}\right) ^{2}\right\}
\nonumber \\
 &+&\frac{2n_+e^{4}}{3m_+^{2}} u_+^0\mathbf{E} \cdot (\mathbf{E}+\mathbf{v_+} \times \mathbf{B})
 \nonumber \\
 &-&  \frac{2n_+e^{4}}{3m_+^{2}\displaystyle{\left( 1-{v_+^{2}}\right) }}u_+^0
\left\{ \left( \mathbf{E}+\mathbf{v_+\times B}\right) ^{2}-%
\left( \mathbf{v_+\cdot E}\right) ^{2}\right\}\label{eq40} \\
\nabla \cdot \mathbf{E} &=& {\rho}_{\rm e}  \\\label{eq41}
\nabla \cdot \mathbf{B} &=& 0   \\\label{eq42}
\frac{\partial\mathbf{B}}{\partial t}  &=& - \nabla \times \mathbf{E} \\\label{eq43}
\mathbf{J} + \frac{\partial\mathbf{E}}{\partial t}  &=& \nabla \times \mathbf{B}\\\label{eq44}
\mathbf{E}+\mathbf{v} \times \mathbf{B}&=&0\\\label{eq45}
\mathbf{u_\pm}&=&\frac{\left  (\rho \gamma \mathbf{v} \pm \frac{m_\mp}{e} \mathbf{J} \right )}{\sqrt{\left  (\rho \gamma \pm \frac{m_\mp}{e} {\rho}_{\rm e} \right )^2- \sum_{n=1}^3\left(\rho \gamma v^n \pm \frac{m_\mp}{e} J^n \right )^2}}\\\label{eq46}
u_\pm^0  &=&  \frac{\left  (\rho \gamma \pm \frac{m_\mp}{e} {\rho}_{\rm e} \right )}{\sqrt{\left  (\rho \gamma \pm \frac{m_\mp}{e} {\rho}_{\rm e} \right )^2- \sum_{n=1}^3\left(\rho \gamma v^n \pm\frac{m_\mp}{e} J^n \right )^2}}\\\label{eq47}
\mathbf{v_\pm}&=&\frac{ \rho \gamma \mathbf{v} \pm \frac{m_\mp}{e} \mathbf{J}  }{  \rho \gamma \pm \frac{m_\mp}{e} {\rho}_{\rm e}  }\\\label{eq48}
n_\pm &=& \frac{\sqrt{\left  (\rho \gamma \pm \frac{m_\mp}{e} {\rho}_{\rm e} \right )^2- \sum_{n=1}^3\left(\rho \gamma v^n \pm \frac{m_\mp}{e} J^n \right )^2}}{m}  \label{eq49}
\end{eqnarray}
where $\mathbf{R}=\gamma^2(e+P)\mathbf{v}+\frac{\mu(e+P)\rho_e\mathbf{J}}{(q)^2}+\mathbf{E\times B}$, $\mathbf{F}=[ ( P+\frac {B^2+E^2} {2} )\mathbf{I}
+\gamma^2(e+P)\mathbf{v}\mathbf{v}+\frac{\mu(e+P)\mathbf{J}\mathbf{J}}{(q)^{2}}
-\mathbf{B}\mathbf{B}-\mathbf{E}\mathbf{E}]$, $\varepsilon=\gamma^2(e+P)+\frac{\mu(e+P)\rho_e^2}{(q)^2}-P+\frac{B^2+E^2}{2}$, $\gamma=\sqrt{\frac{1}{1-((\frac{u^1}{u^0})^2+(\frac{u^2}{u^0})^2+(\frac{u^3}{u^0})^2)}}$ is the Lorentz factor and $e=\rho+\frac{P}{\Gamma-1}$, $\Gamma$ is the specific heat ratio.

To get a simplified set of the above equations, we substitute equations (\ref{eq45})--(\ref{eq49}) into equations (\ref{39})--(\ref{eq40}), and we get the equations in Appendix A. Noting that $m_+=\frac{m\Delta \mu +\sqrt{m^2\Delta\mu^2+4m^2\mu}}{2}$ and $m_-=\frac{-m\Delta \mu +\sqrt{m^2\Delta\mu^2+4m^2\mu}}{2}$, where $m=m_++m_-$, $\mu=\frac{m_+m_-}{m^2}$ is the normalized reduced mass and $\Delta\mu=\frac{m_+-m_-}{m}$ is the normalized mass difference, when we substitute these into equations (\ref{eq11152})--(\ref{eq11153}) in Appendix A, we could get the equations with $m$, $\mu$, $\Delta \mu$ instead of $m_+$ and $m_-$ in Appendix B.

The one-fluid relativistic magnetohydrodynamics description of two-fluid plasmas with Landau-Lifshitz radiation reaction has been complished. The one-fluid description of such two-fluid plasmas with Landau-Lifshitz radiation reaction in curved spacetime can also be achieved through the same method described above.

To compare the Lorentz force with the Landau-Lifshitz radiation reaction force acting on the fluid element in extreme astrophysical situations, jet along magnetic field line in Gamma-ray burst for example, we first get the Landau-Lifshitz radiation reaction force from the right side of equation (\ref{eq11152}) in Appendix A with the condition that the relativistic ion-electron plasma is neutral and the flow is parallel to the magnetic field line which is perpendicular to the current density for simplicity (see Appendix C for the detailed derivation. Here, for the sake of quantitative analysis, we conduct the derivation without setting $c=1$)
\begin{eqnarray}
\mathbf{F}_{LL}  &=&  \frac{2}{3} r_e^2\left[\left( \mathbf{B}\times\left(\frac{\mathbf{J}}{ce}\times\mathbf{B}\right)\right)+(\frac{\mathbf{v}}{c}\times\mathbf{B})(\frac{\mathbf{v}}{c}\cdot(\frac{\mathbf{J}}{ce}\times\mathbf{B}))\right]
-\frac{2}{3} n r_e^2\gamma^2[(\frac{{\mathbf{J}}\times\mathbf{B}}{cne})^2-(\frac{{\mathbf{J}}\times\mathbf{B}}{cne}\cdot\frac{\mathbf{v}}{c})^2](\frac{\mathbf{v}}{c}-\frac{\mathbf{J}}{cne}) \label{eq50}
\end{eqnarray}
where $r_e=\frac{e^2}{m_e c^2}$, $n=\frac{\gamma\rho}{m} \approx \frac{\gamma\rho}{m_+}$ for ion-electron plasma and c is the speed of light.

In ultra-relativistic conditions, the second term is larger than the first due to $\gamma^2$ in equation (\ref{eq50}), and we find that the first term can be reduced to

\begin{equation}
\frac{2}{3} \frac{|\mathbf{B}|}{B_c}|\left(\frac{\mathbf{J}\times\mathbf{B}}{c}\right)| \label{eq51}
\end{equation}
where $B_c=m^2_{e}c^4/e^3$

Then the ratio of formula (\ref{eq51}) to Lorentz force is $\frac{2B}{3B_c}$. The value of second term is reduced to
\begin{equation}
 \frac{2}{3} n r_e^2\gamma^2(\frac{{\mathbf{J}}\times\mathbf{B}}{cne})^2 \label{eq52}
\end{equation}

then the ratio of it to Lorentz force is
\begin{equation}
\frac{2}{3} \frac{r_e^2}{e}\gamma^2 \frac{|\mathbf{J}||\mathbf{B}|}{cne}\sim  \frac{2}{3} \frac{r_e^2}{e}\gamma^2 |\mathbf{B}| \frac{|\mathbf{v-\frac{\gamma_e}{\gamma} v_e}|}{c} \sim \frac{2}{3} \frac{B}{B_c} \gamma^2 \frac{|\mathbf{v-\frac{\gamma_e}{\gamma} v_e}|}{c}   \label{eq53}
\end{equation}
where we assume $\gamma \approx \gamma_+$ and the current density is perpendicular to the magnetic field for simplicity, $\mathbf{v_e}$ and $\gamma_e$ are the velocity and Lorentz factor of electron in fluid element, respectively.

We find from equation (\ref{eq53}) that if $\mathbf{v-\frac{\gamma_e}{\gamma} v_e}$ in the fluid element is mildly relativistic, the radiation reaction force can be comparable to Lorentz force in the condition in which $B=10^{10}$ G and $\gamma=10^3$ for example.

In work by \cite{19} in which the radiation reaction force is included to investigate particle acceleration in ultra-relativistic pair plasma reconnection using particle-in-cell method, with the initial $\mathbf{E_0}=0$, $B_0=5mG$ and the background particles with Lorentz factor from $\gamma=4\times 10^7$ to $\gamma=4\times 10^8$ with a power-law index of $2$, we get the value of Landau-Lifshitz radiation reaction force acting on the positive and negative particles in the fluid element $\frac{2}{3} n_{\pm} r_e^2 \gamma^2 B_0^2$. With the Lorentz force whose value is $n_{\pm}eB_0$ for the positive and negative particles in the fluid element, then we get the ratio of radiation reaction force to Lorentz force $\frac{2}{3} \frac{B_0}{B_c} \gamma^2 \sim 10^{-2}$ for both positive and negative particles from the fluid element when taking $\gamma=4\times10^8$. During the pair plasma evolution in \cite{19}, the Lorentz factor of the background particle can reach $10^9$ which may show that radiation reaction force acting on positive or negative particles in the fluid element can be comparable to Lorentz force acting on it.

By considering the relativistic fluid in the rest frame, we investigate waves propagating in a rest plasma embedded in a uniform magnetic field $\bm{B_0} = (0, 0, B_0)$. Then we linearize these equations by separating the variables into background fields with subscripts 0 and perturbations with tilde symbols proportional to $\exp(i\bm{k}\cdot\bm{x} - \omega t)$ where the wave vector $\bm{k} = (k_\perp,\, 0,\, 0)$ and $\omega$ is the frequency. The background conditions lead $\bm{v_0} = 0, \rho_{e0} = 0, \; \bm{E_0} = 0$, and $\bm{J_0} = 0$, and we assume the plasma is neutral everywhere. Then, the linearized equations of perturbations are given by

\begin{eqnarray}
\frac{\partial}{\partial t} \tilde{\rho}+ \rho_0 \nabla \cdot \tilde{\VEC{v}} &=& 0   \label{eq54} \\
h_0 \frac{\partial}{\partial t} \tilde{\VEC{v}}  &=& - c^2\nabla \tilde{p}+ c\tilde{\VEC{J}} \times \VEC{B}_0
+\frac{2}{3}\frac{r_e^2c}{e}\VEC{B}_0\times(\tilde{\VEC{J}} \times \VEC{B}_0) \label{eq55} \\
\nabla \cdot \tilde{\VEC{B}} &=& 0  \label{eq56} \\
\frac{\partial}{ \partial t} \tilde{\VEC{B}} &=& - c\nabla \times \tilde{\VEC{E}} \label{eq57}\\
\tilde{\VEC{J}} + \frac{\partial}{ \partial t} \tilde{\VEC{E}}&=& c\nabla \times \tilde{\VEC{B}}\label{eq58} \\
(\rho_0/p_0) (\tilde{p}/\tilde{\rho}) &=& \Gamma\label{eq59}\\
\tilde{\bm{E}} + \f{\tilde{\bm{v}}}{c} \times\bm{B}_0 &=& 0\label{eq60}
\end{eqnarray}
where we set $\Gamma=\frac{4}{3}$

Substituting perturbations proportional to $\exp(i\bm{k}\cdot\bm{x} - \omega t)$ into above equations, we get
\begin{eqnarray}
  -i\omega \tilde{\rho} + i \rho_0 \bm{k}\cdot\tilde{\bm{v}} &=& 0\label{eq61} \\
  -i\omega h_0\tilde{\bm{v}} &=& -i\bm{k}c^2\tilde{p} + c\tilde{\bm{J}}\times\bm{B}_0
+\frac{2}{3}\frac{r_e^2c}{e}\VEC{B}_0\times(\tilde{\VEC{J}} \times \VEC{B}_0)\label{eq62} \\
    i\bm{k}\cdot\tilde{\bm{B}} &=& 0\label{eq66} \\
  i\bm{k}\times\tilde{\bm{E}} &=& \f{1}{c}i\omega\tilde{\bm{B}}\label{eq67}\\
    \f{\tilde{\bm{J}}}{c} &=& i\bm{k}\times\tilde{\bm{B}} + \f{1}{c}i\omega\tilde{\bm{E}} \label{eq65}\\
  -i\omega\tilde{p} &=& -\f{\Gamma p_0}{\rho_0}i\omega\tilde{\rho} \label{eq64}\\
  \tilde{\bm{E}} + \f{\tilde{\bm{v}}}{c} \times\bm{B}_0 &=& 0 \label{eq63}
\end{eqnarray}

By $\VEC{k}\times$ (\ref{eq63}) combing with equation (\ref{eq67}), we get $\tilde{\VEC{B}}$. Substituting $\tilde{\VEC{E}}$ from equation (\ref{eq63}) and $\tilde{\VEC{B}}$ into equation (\ref{eq65}) yields $\tilde{\VEC{J}}$. By substituting $\tilde{\rho}$ from equation (\ref{eq61}) into equation (\ref{eq64}), we get $\tilde{p}$. Then substituting $\tilde{\VEC{J}}$ and $\tilde{p}$ into equation (\ref{eq62}), we get the dispersion relation

\begin{equation}
\omega^2=\frac{(\Gamma p_0 c^2 + B_0^2 c^2)(h_0+B_0^2)+(\frac{2}{3}\frac{r_e^2}{e})^2 B_0^6 c^2}{(h_0+B_0^2)^2 +(\frac{2}{3}\frac{r_e^2}{e})^2 B_0^6 } k_\perp^2
\end{equation}

the normalized phase velocity
\begin{equation}
(v_{ph})^2=(\frac{\omega}{c k_\perp})^2=\frac{(\Gamma p_0 + B_0^2)(h_0+B_0^2)+(\frac{2}{3}\frac{r_e^2}{e})^2 B_0^6}{(h_0+B_0^2)^2 +(\frac{2}{3}\frac{r_e^2}{e})^2 B_0^6 }
\end{equation}

It is noted that $\frac{\Gamma p_0 + B_0^2}{h_0+B_0^2}<1$, then we find that $\frac{(\Gamma p_0 + B_0^2)(h_0+B_0^2)+(\frac{2}{3}\frac{r_e^2}{e})^2 B_0^6}{(h_0+B_0^2)^2 +(\frac{2}{3}\frac{r_e^2}{e})^2 B_0^6 }>\frac{\Gamma p_0 + B_0^2}{h_0+B_0^2}$, meaning that the normalized phase velocity with Landau-Lifshitz radiation reaction is larger than that without Landau-Lifshitz radiation reaction.

\section{Discussion}
The one-fluid relativistic magnetohydrodynamics equations using a relativistic two-fluid approximation of plasma consisting
of positively and negatively charged particles in which the Landau-Lifshitz radiation reaction force is included, provide a detailed description containing a self-consistent expression of the Landau-Lifshitz radiation reaction force acting on the charged particles in the plasma. The equations  derived in this work is a natural generalization of modern relativistic magnetohydrodynamics and could be used to many situations in relativistic plasmas.

The radiation reaction force is neglected in most of previous studies due to its small value compared to the Lorentz force, thus, being not
expected to be a major effect on the dynamics of plasma. This
assumption may not be so suitable for situations in some astrophysical conditions
where spatiotemporal scales of plasma's motion are so sufficiently
large that the effect of radiation reaction couldn't be neglectable,like the work by \cite{19} in which the radiation reaction force is included to investigate particle acceleration in ultra-relativistic pair plasma reconnection using particle-in-cell method. In astrophysical conditions like Gamma-ray burst, the inner region of accretion disk around neutron star or black hole, the inner region of active galactic nuclei where plasma motion is relativistic and the background magnetic field is large, the radiation reaction may not be negligible, then the one-fluid relativistic magnetohydrodynamics description of two-fluid plasmas with Landau-Lifshitz radiation reaction derived in this work could be applied to these astrophysical conditions for simulations. Here, from the qualitative point of view, we may find some distinctive phenomenon from the extreme astrophysical conditions when using the equations in this work. Considering an electron moving in a stable orbit around a black hole immersed in a large magnetic field perpendicular to the orbital plane of the electron, due to the radiation reaction, the kinetic energy of the electron is decreased. Depending on the orientation of
the Lorentz force on the electron with respect to the black hole, the radiation reaction will lead to the fall of the charged particle into the black hole when the direction of the Lorentz force is toward the black hole,otherwise the orbit of the electron will remain bounded while oscillations is decaying when Lorentz force is outward the black hole \citep{49}. When considering accretion disk around such black hole, we may expect that accretion rate of material flowing into black hole with radiation reaction is different from that without radiation reaction. The numerical simulation using equations derived in this work applied to these astrophysical conditions will be conducted in future.

\section*{Acknowledgements}
We are very grateful to the anonymous referee for her/his instructive
comments which improved the content of the paper. This work is supported by the National Key Research and Development Program of China (No. 2017YFA0402703). This work
has been supported by the National Science Foundations of China
(Nos. 11373024, 11173016 and 11233003).

\appendix

\section*{a}
 When we substitute equations (\ref{eq45})--(\ref{eq49}) into equations (\ref{39})--(\ref{eq40}), we get the following relativistic magnetohydrodynamic equations containing Landau-Lifshitz radiation reaction with $m_+$ and $m_-$
\begin{eqnarray}
\frac{\partial (\gamma{\rho})}{\partial t}+\nabla \cdot (\gamma\rho\mathbf{v}) &=& 0 \label{eq1151}\\
\frac{\partial \mathbf{R}}{\partial t}
+\nabla \cdot \mathbf{F} &=&
\frac{2e^{4}}{3m_-^{2}}\frac{\left  (\rho \gamma - \frac{m_+}{e} {\rho}_{\rm e} \right )}{m}\left\{ \mathbf{E\times B}
+
\left( \mathbf{B}\times \left( \mathbf{B}\times(\frac{  \rho \gamma \mathbf{v} - \frac{m_+}{e} \mathbf{J}  }{  \rho \gamma - \frac{m_+}{e} {\rho}_{\rm e}  })\right)\right)
+
\mathbf{E}\left( (\frac{  \rho \gamma \mathbf{v} - \frac{m_+}{e} \mathbf{J}  }{  \rho \gamma - \frac{m_+}{e} {\rho}_{\rm e}  })\cdot \mathbf{E}\right) \right\}
\nonumber \\
&-&  \frac{2e^{4}}{3m_-^{2}\displaystyle{\left( 1-{(\frac{  \rho \gamma \mathbf{v} - \frac{m_+}{e} \mathbf{J}  }{  \rho \gamma - \frac{m_+}{e} {\rho}_{\rm e}  })^{2}}\right) }}
\frac{\left  (\rho \gamma \mathbf{v} - \frac{m_+}{e} \mathbf{J} \right )}{m}\left\{ \left( \mathbf{E}
+
(\frac{ \rho \gamma \mathbf{v} - \frac{m_+}{e} \mathbf{J}  }{  \rho \gamma - \frac{m_+}{e} {\rho}_{\rm e}  })\times \mathbf{B}\right) ^{2}-
\left( (\frac{  \rho \gamma \mathbf{v} - \frac{m_+}{e} \mathbf{J}  }{  \rho \gamma - \frac{m_+}{e} {\rho}_{\rm e}  })\cdot \mathbf{E}\right) ^{2}\right\}
\nonumber \\
&+&\frac{2e^{4}}{3m_+^{2}}\frac{\left  (\rho \gamma + \frac{m_-}{e} {\rho}_{\rm e} \right )}{m}\left\{ \mathbf{E\times B}
+
\left( \mathbf{B}\times \left( \mathbf{B}\times(\frac{  \rho \gamma \mathbf{v} + \frac{m_-}{e} \mathbf{J}  }{  \rho \gamma + \frac{m_-}{e} {\rho}_{\rm e}  })\right)\right)
+
\mathbf{E}\left( (\frac{  \rho \gamma \mathbf{v} + \frac{m_-}{e} \mathbf{J}  }{  \rho \gamma + \frac{m_-}{e} {\rho}_{\rm e}  })\cdot \mathbf{E}\right) \right\}
\nonumber \\
&-&  \frac{2e^{4}}{3m_+^{2}\displaystyle{\left( 1-{(\frac{  \rho \gamma \mathbf{v} + \frac{m_-}{e} \mathbf{J}  }{  \rho \gamma + \frac{m_-}{e} {\rho}_{\rm e}  })^{2}}\right) }}
\frac{\left  (\rho \gamma \mathbf{v} + \frac{m_-}{e} \mathbf{J} \right )}{m}\left\{ \left( \mathbf{E}
 +
(\frac{ \rho \gamma \mathbf{v} + \frac{m_-}{e} \mathbf{J}  }{  \rho \gamma + \frac{m_-}{e} {\rho}_{\rm e}  })\times \mathbf{B}\right) ^{2}-
\left( (\frac{  \rho \gamma \mathbf{v} + \frac{m_-}{e} \mathbf{J}  }{  \rho \gamma + \frac{m_-}{e} {\rho}_{\rm e}  })\cdot \mathbf{E}\right) ^{2}\right\}    \label{eq11152}\\
\frac{\partial \varepsilon}{\partial t}+\nabla \cdot \mathbf{R} &=&
\frac{2e^{4}}{3m_-^{2}} \frac{\left  (\rho \gamma - \frac{m_+}{e} {\rho}_{\rm e} \right )}{m}\mathbf{E} \cdot (\mathbf{E}+(\frac{  \rho \gamma \mathbf{v} - \frac{m_+}{e} \mathbf{J}  }{  \rho \gamma - \frac{m_+}{e} {\rho}_{\rm e}  }) \times \mathbf{B})
 \nonumber \\
 &-&  \frac{2e^{4}}{3m_-^{2}\displaystyle{\left( 1-{(\frac{  \rho \gamma \mathbf{v} - \frac{m_+}{e} \mathbf{J}  }{  \rho \gamma - \frac{m_+}{e} {\rho}_{\rm e}  })^{2}}\right) }}
 \frac{\left  (\rho \gamma - \frac{m_+}{e} {\rho}_{\rm e} \right )}{m}
\left\{ \left( \mathbf{E}+(\frac{  \rho \gamma \mathbf{v} - \frac{m_+}{e} \mathbf{J}  }{  \rho \gamma - \frac{m_+}{e} {\rho}_{\rm e}  })\times \mathbf{B}\right) ^{2}%
-\left( (\frac{  \rho \gamma \mathbf{v} - \frac{m_+}{e} \mathbf{J}  }{  \rho \gamma - \frac{m_+}{e} {\rho}_{\rm e}  })\cdot \mathbf{E}\right) ^{2}\right\}
\nonumber \\
&+&\frac{2e^{4}}{3m_+^{2}} \frac{\left  (\rho \gamma + \frac{m_-}{e} {\rho}_{\rm e} \right )}{m}\mathbf{E} \cdot (\mathbf{E}+(\frac{  \rho \gamma \mathbf{v} + \frac{m_-}{e} \mathbf{J}  }{  \rho \gamma + \frac{m_-}{e} {\rho}_{\rm e}  }) \times \mathbf{B})
  \nonumber \\
 &-&  \frac{2e^{4}}{3m_+^{2}\displaystyle{\left( 1-{(\frac{  \rho \gamma \mathbf{v} + \frac{m_-}{e} \mathbf{J}  }{  \rho \gamma + \frac{m_-}{e} {\rho}_{\rm e}  })^{2}}\right) }}
\frac{\left  (\rho \gamma + \frac{m_-}{e} {\rho}_{\rm e} \right )}{m}
\left\{ \left( \mathbf{E}+(\frac{  \rho \gamma \mathbf{v} + \frac{m_-}{e} \mathbf{J}  }{  \rho \gamma + \frac{m_-}{e} {\rho}_{\rm e}  })\times \mathbf{B}\right) ^{2}%
-\left( (\frac{  \rho \gamma \mathbf{v} + \frac{m_-}{e} \mathbf{J}  }{  \rho \gamma + \frac{m_-}{e} {\rho}_{\rm e}  })\cdot \mathbf{E}\right) ^{2}\right\} \label{eq11153} \\
\nabla \cdot \mathbf{E} &=& {\rho}_{\rm e}  \\\label{eq1154}
\nabla \cdot \mathbf{B} &=& 0   \\\label{eq1155}
\frac{\partial\mathbf{B}}{\partial t}  &=& - \nabla \times \mathbf{E} \\\label{eq1156}
\mathbf{J} + \frac{\partial\mathbf{E}}{\partial t}  &=& \nabla \times \mathbf{B}\\\label{eq1157}
\mathbf{E}+\mathbf{v} \times \mathbf{B}&=&0\label{eq1158}
\end{eqnarray}
where $m_+=\frac{m\Delta \mu +\sqrt{m^2\Delta\mu^2+4m^2\mu}}{2}$ and $m_-=\frac{-m\Delta \mu +\sqrt{m^2\Delta\mu^2+4m^2\mu}}{2}$

\section*{b}
 When we substitute $m_+=\frac{m\Delta \mu +\sqrt{m^2\Delta\mu^2+4m^2\mu}}{2}$ and $m_-=\frac{-m\Delta \mu +\sqrt{m^2\Delta\mu^2+4m^2\mu}}{2}$ into equations (\ref{eq11152})--(\ref{eq11153}) in Appendix A with the relation $m=m_++m_-$, $\mu=\frac{m_+m_-}{m^2}$ and $\Delta\mu=\frac{m_+-m_-}{m}$, then we get the following relativistic magnetohydrodynamic equations containing Landau-Lifshitz radiation reaction with $m$, $\mu$ and $\Delta\mu$
\begin{eqnarray*}
\frac{\partial (\gamma{\rho})}{\partial t}+\nabla \cdot (\gamma\rho\mathbf{v}) &=& 0 \\
\frac{\partial \mathbf{R}}{\partial t}
+\nabla \cdot \mathbf{F}&=&\frac{2e^4}{3m}\{\frac{(1-2\mu)\rho\gamma}{m^2\mu^2}[\mathbf{E}\times\mathbf{B}+\mathbf{B}\times(\mathbf{B}\times\mathbf{v})+(\mathbf{v}\cdot\mathbf{E})\mathbf{E}]
-\frac{\Delta\mu(1-\mu)}{m\mu^2e}[\rho_e(\mathbf{E}\times\mathbf{B})+\mathbf{B}\times(\mathbf{B}\times\mathbf{J})+(\mathbf{J}\cdot\mathbf{E})\mathbf{E}]\}
\nonumber \\
 &-&\frac{2e^4}{3m^3\mu^2}\{(a_1 b_1+a_2 b_2m^4\mu^2-a_3 b_3m^2\mu)[\rho\gamma\mathbf{v}(1-2\mu)-\frac{\mathbf{J}}{e}(\Delta\mu m(1-\mu))]
\nonumber \\
&+&(a_3b_2m^2\mu-a_1b_3)[\rho\gamma\mathbf{v}(\Delta\mu m(1-\mu))-\frac{\mathbf{J}}{e}((m-2m\mu)^2-2m^2\mu^2)]
\nonumber \\
&+&(a_3b_1\mu-a_2b_3m^2\mu^2)[\rho\gamma\mathbf{v}\Delta \mu m - \frac{\mathbf{J}}{e}(m^2-2m^2\mu)]
\nonumber \\
&+&a_1b_2[\rho\gamma\mathbf{v}((m-2m\mu)^2-2m^2\mu^2)-\frac{\mathbf{J}}{e}(m\Delta \mu((m-2m\mu)^2-2m^2\mu^2+\mu(m^2-2m^2\mu)+m^2\mu^2))]
\nonumber \\
&+&a_2b_1m^2\mu^2[2\rho\gamma\mathbf{v}-\frac{\mathbf{J}}{e}m\Delta \mu]\}/
[a_1^2+a_1a_2(m^2-2m^2\mu)-a_1a_3m\Delta \mu +a_2a_3m^3\mu\Delta\mu+a_2^2m^4\mu^2-a_3^2m^2\mu]\\
\frac{\partial \varepsilon}{\partial t}+\nabla \cdot \mathbf{R}&=&\frac{2e^4}{3m^3\mu^2}\mathbf{E}\cdot[\rho\gamma(\mathbf{E}+\mathbf{v}\times\mathbf{B})(1-2\mu)-\frac{\rho_e\mathbf{E}+\mathbf{J}\times\mathbf{B}}{e}(\Delta \mu m(1-\mu))]
\nonumber \\
&-&\frac{2e^4}{3m^3\mu^2}\{(a_1 b_1+a_2 b_2m^4\mu^2-a_3 b_3m^2\mu)[\rho\gamma(1-2\mu)-\frac{\rho_e}{e}(\Delta\mu m(1-\mu))]
\nonumber \\
&+&(a_3b_2m^2\mu-a_1b_3)[\rho\gamma(\Delta\mu m(1-\mu))-\frac{\rho_e}{e}((m-2m\mu)^2-2m^2\mu^2)]
\nonumber \\
&+&(a_3b_1\mu-a_2b_3m^2\mu^2)[\rho\gamma\Delta \mu m - \frac{\rho_e}{e}(m^2-2m^2\mu)]
\nonumber \\
&+&a_1b_2[\rho\gamma((m-2m\mu)^2-2m^2\mu^2)-\frac{\rho_e}{e}(m\Delta \mu((m-2m\mu)^2-2m^2\mu^2+\mu(m^2-2m^2\mu)+m^2\mu^2))]
\nonumber \\
&+&a_2b_1m^2\mu^2[2\rho\gamma-\frac{\rho_e}{e}m\Delta \mu]\}/
[a_1^2+a_1a_2(m^2-2m^2\mu)-a_1a_3m\Delta \mu +a_2a_3m^3\mu\Delta\mu+a_2^2m^4\mu^2-a_3^2m^2\mu]\\
\nabla \cdot \mathbf{E} &=& {\rho}_{\rm e}  \\
\nabla \cdot \mathbf{B} &=& 0   \\
\frac{\partial\mathbf{B}}{\partial t}  &=& - \nabla \times \mathbf{E} \\
\mathbf{J} + \frac{\partial\mathbf{E}}{\partial t}  &=& \nabla \times \mathbf{B}\\
\mathbf{E}+\mathbf{v} \times \mathbf{B}&=&0
\end{eqnarray*}

where

\begin{eqnarray*}
a_1&=&\rho^2\gamma^2-\rho^2\gamma^2 v^2\\
a_2&=&\frac{\rho_e^2-J^2}{e^2}\\
a_3&=&\frac{2\rho\gamma\rho_e-2\rho\gamma\mathbf{v}\cdot\mathbf{J}}{e}\\
b_1&=&\rho^2\gamma^2(\mathbf{E}+\mathbf{v}\times\mathbf{B})^2-\rho^2\gamma^2(\mathbf{v}\cdot\mathbf{E})^2\\
b_2&=&\frac{(\rho_e\mathbf{E}+\mathbf{J}\times\mathbf{B})^2-(\mathbf{J}\cdot\mathbf{E})^2}{e^2}\\
b_3&=&\frac{2\rho\gamma(\mathbf{E}+\mathbf{v}\times\mathbf{B})\cdot(\rho_e\mathbf{E}+\mathbf{J}\times\mathbf{B})-2\rho\gamma(\mathbf{v}\cdot\mathbf{E})(\mathbf{J}\cdot\mathbf{E})}{e}
\end{eqnarray*}

\section*{c}
The Landau-Lifshitz radiation reaction force from the right side of equation (\ref{eq11152}) in Appendix A acting on ion-electron fluid element is (the radiation reaction force acting on ion is neglected due to $m_i \gg m_e$. Here, we conduct the derivation without setting $c=1$)
\begin{eqnarray*}
\mathbf{F_{LL}} &= &
\frac{2e^{4}}{3m_-^{2}c^4}\frac{\left  (\rho \gamma - \frac{m_+}{e} {\rho}_{\rm e} \right )}{m}\left\{ \mathbf{E\times B}
 +
\frac{1}{c}\left( \mathbf{B}\times \left( \mathbf{B}\times(\frac{  \rho \gamma \mathbf{v} - \frac{m_+}{e} \mathbf{J}  }{  \rho \gamma - \frac{m_+}{e} {\rho}_{\rm e}  })\right)\right)
+
\frac{1}{c}\mathbf{E}\left( (\frac{  \rho \gamma \mathbf{v} - \frac{m_+}{e} \mathbf{J}  }{  \rho \gamma - \frac{m_+}{e} {\rho}_{\rm e}  })\cdot \mathbf{E}\right) \right\}
\nonumber \\
&-&  \frac{2e^{4}}{3m_-^{2}c^5\displaystyle{\left( 1-\frac{1}{c^2}{(\frac{  \rho \gamma \mathbf{v} - \frac{m_+}{e} \mathbf{J}  }{  \rho \gamma - \frac{m_+}{e} {\rho}_{\rm e}  })^{2}}\right) }}
\frac{\left  (\rho \gamma \mathbf{v} - \frac{m_+}{e} \mathbf{J} \right )}{m}\left\{ \left( \mathbf{E}
 +
(\frac{1}{c}\frac{ \rho \gamma \mathbf{v} - \frac{m_+}{e} \mathbf{J}  }{  \rho \gamma - \frac{m_+}{e} {\rho}_{\rm e}  })\times \mathbf{B}\right) ^{2}-
\frac{1}{c^2}\left( (\frac{  \rho \gamma \mathbf{v} - \frac{m_+}{e} \mathbf{J}  }{  \rho \gamma - \frac{m_+}{e} {\rho}_{\rm e}  })\cdot \mathbf{E}\right) ^{2}\right\} \label{eq151}\\
\end{eqnarray*}

When we assume the plasma is neutral, ${\rho}_{\rm e}=0$, then the above equation changes to
\begin{eqnarray*}
\mathbf{F_{LL}}& =&
\frac{2e^{4}}{3m_-^{2}c^4}\frac{\rho \gamma }{m}\left\{ \mathbf{E\times B}
 +
\frac{1}{c}\left( \mathbf{B}\times \left( \mathbf{B}\times\frac{  \rho \gamma \mathbf{v} - \frac{m_+}{e} \mathbf{J}  }{  \rho \gamma   }\right)\right)
+
\frac{1}{c}\mathbf{E}\left( \frac{  \rho \gamma \mathbf{v} - \frac{m_+}{e} \mathbf{J}  }{  \rho \gamma   }\cdot \mathbf{E}\right) \right\}
\nonumber \\
&-&  \frac{2e^{4}}{3m_-^{2}c^5\displaystyle{\left( 1-\frac{1}{c^2}{(\frac{  \rho \gamma \mathbf{v} - \frac{m_+}{e} \mathbf{J}  }{  \rho \gamma   })^{2}}\right) }}
\frac{\left  (\rho \gamma \mathbf{v} - \frac{m_+}{e} \mathbf{J} \right )}{m}\left\{ \left( \mathbf{E}
 +
\frac{1}{c}\frac{ \rho \gamma \mathbf{v} - \frac{m_+}{e} \mathbf{J}  }{  \rho \gamma   }\times \mathbf{B}\right) ^{2}-
\frac{1}{c^2}\left( \frac{  \rho \gamma \mathbf{v} - \frac{m_+}{e} \mathbf{J}  }{  \rho \gamma   }\cdot \mathbf{E}\right) ^{2}\right\} \label{eq151}\\
\end{eqnarray*}

In ideal relativistic magnetohydrodynamic, $\mathbf{E}=-\frac{1}{c}\mathbf{v} \times \mathbf{B}$. Substituting this equation into above, we get

\begin{eqnarray*}
\mathbf{F}_{LL}  =  \frac{2}{3} r_e^2\left[\left( \mathbf{B}\times\left(\frac{\mathbf{J}}{ce}\times\mathbf{B}\right)\right)+(\frac{\mathbf{v}}{c}\times\mathbf{B})(\frac{\mathbf{v}}{c}\cdot(\frac{\mathbf{J}}{ce}\times\mathbf{B}))\right]
-\frac{2}{3} n r_e^2\gamma^2\frac{1}{\displaystyle{\left( 1-\frac{1}{c^2}{(  \mathbf{v} - \frac{\mathbf{J}}{ne}   )}^{2}\right)} }[(\frac{{\mathbf{J}}\times\mathbf{B}}{cne})^2-(\frac{{\mathbf{J}}\times\mathbf{B}}{cne}\cdot\frac{\mathbf{v}}{c})^2](\frac{\mathbf{v}}{c}-\frac{\mathbf{J}}{cne})
\end{eqnarray*}
where $r_e=\frac{e^2}{m_e c^2}$, $n=\frac{\gamma\rho}{m} \approx \frac{\gamma\rho}{m_+}$ for ion-electron plasma and c is the speed of light.

When assuming the current density is perpendicular to the magnetic field which is parallel the velocity of the plasma, we get $\frac{1}{\displaystyle{( 1-\frac{1}{c^2}{  \mathbf{v}    }^{2})}} < \frac{1}{\displaystyle{( 1-\frac{1}{c^2}{(  \mathbf{v} - \frac{\mathbf{J}}{ne}   )}^{2})}} $, that is to say, $\gamma^2 < \gamma_-^2$. Here we take $\gamma_-^2 \approx \gamma^2$, then the above equation becomes
\begin{eqnarray*}
\mathbf{F}_{LL}  =\frac{2}{3} r_e^2\left[\left( \mathbf{B}\times\left(\frac{\mathbf{J}}{ce}\times\mathbf{B}\right)\right)+(\frac{\mathbf{v}}{c}\times\mathbf{B})(\frac{\mathbf{v}}{c}\cdot(\frac{\mathbf{J}}{ce}\times\mathbf{B}))\right]
-\frac{2}{3} n r_e^2\gamma^2\frac{1}{\displaystyle{\left( 1-\frac{1}{c^2}{  \mathbf{v}}^{2}\right)} }[(\frac{{\mathbf{J}}\times\mathbf{B}}{cne})^2-(\frac{{\mathbf{J}}\times\mathbf{B}}{cne}\cdot\frac{\mathbf{v}}{c})^2](\frac{\mathbf{v}}{c}-\frac{\mathbf{J}}{cne})
\end{eqnarray*}



\bibliographystyle{apj}

\end{document}